\documentclass[%
reprint,
superscriptaddress,
showpacs,preprintnumbers,
nofootinbib,
amsmath,amssymb,
prx
]{revtex4-1}

\usepackage{graphicx}
\usepackage{dcolumn}
\usepackage{bm}

\usepackage{amssymb}
\usepackage{amsmath}
\usepackage{amsfonts}
\usepackage{xspace}
\usepackage{bm,bbm}
\usepackage{relsize}
\usepackage{siunitx}
\usepackage[usenames,dvipsnames]{color}

\renewcommand{\vec}[1]{\ensuremath{\bm{#1}}}

\newcommand{\ie}{\emph{i.e.}\xspace}
\newcommand{\cf}{\emph{cf.}\xspace}

\newcommand{\identity}{\ensuremath{\mathlarger{\mathbbm{1}}}}
\renewcommand{\eqref}[1]{Eq.~\ref{eq:#1}}
\newcommand{\figref}[1]{Fig.~\ref{fig:#1}}
\newcommand{\tabref}[1]{Tab.~\ref{tab:#1}}

\newcommand*{\cc}[1]{\ensuremath{\,\overline{#1\vphantom{\bar{#1}}}\,}}
\newcommand{\eq}{\ensuremath{\,{=}\,}}
\newcommand{\defas}{\ensuremath{\,{:=}\,}}
\newcommand{\asdef}{\ensuremath{\,{=:}\,}}

\begin{document}

\title{The Nature of Topological Protection in Spin and Valley Hall Insulators} 

\author{Matthias Saba}
\email[Email: ]{m.saba@imperial.ac.uk}
\affiliation{The Blackett Laboratory, Imperial College London, London SW7 2AZ, UK}
\affiliation{Adolphe Merkle Institute, University of Fribourg, 1700 Fribourg, Switzerland}
\author{Stephan Wong}
\affiliation{The Blackett Laboratory, Imperial College London, London SW7 2AZ, UK}
\affiliation{School of Physics and Astronomy, Cardiff University, Cardiff CF24 3AA, UK}
\author{Mathew Elman}
\affiliation{The Blackett Laboratory, Imperial College London, London SW7 2AZ, UK}
\author{Sang Soon Oh}
\affiliation{School of Physics and Astronomy, Cardiff University, Cardiff CF24 3AA, UK}
\author{Ortwin Hess}
\email[Email: ]{o.hess@imperial.ac.uk}
\affiliation{The Blackett Laboratory, Imperial College London, London SW7 2AZ, UK}

\date{\today}

\begin{abstract}
    Recent interest in optical analogues to the quantum spin Hall and quantum valley Hall effects is driven by the promise to establish topologically protected photonic edge modes at telecommunication and optical wavelengths on a simple platform suitable for industrial applications. While first theoretical and experimental efforts have been made, these approaches so far both lack a rigorous understanding of the nature of topological protection and the limits of backscattering immunity. We here use a generic group theoretical methodology to fill this gap and obtain general design principles for purely dielectric two-dimensional topological photonic systems. The method comprehensively characterizes possible 2D hexagonal designs and reveals their topological nature, potential and limits.
\end{abstract}

\maketitle

Since Haldane and Raghu \cite{PhysRevLett.100.013904,PhysRevA.78.033834} proposed the optical analogue of the integer quantum Hall effect in a 2D photonic crystal (PhC) using gyromagnetic materials in 2008, the field of topological photonics has been growing. The broad interest in topological photonic systems mainly stems from the promise of unidirectional, backscattering-free interface waves that are protected against material and fabrication imperfections, and environmental fluctuations \cite{Lu2014,Lu2016,khanikaev2017two}. These properties bear the potential to solve many of the state of the art problems connected with on-chip optical computation and data processing.

The original proposal quickly sparked subsequent experimental studies using gyromagnetic and gyroelectric effects \cite{Wang2009}. The need of extremely large magnetic fields to observe the predicted behaviour at higher frequencies, however, limits the scope of applications. A number of workarounds to reach higher frequencies without the need of gigantic static magnetic fields include emulating reciprocity breaking through coupled resonators \cite{fang2012realizing}, and engineering of anti-symmetric scattering matrices in parity-time-duality symmetric systems \cite{PhysRevB.95.035153,khanikaev2013photonic}. These approaches, however, either require structures that are much larger than the operation wavelength or exotic constituent materials.

Designs based on purely dielectric platforms \cite{PhysRevLett.114.223901,chen2017valley,ma2016all}, conceivably related to the Quantum Spin Hall effect (QSHE) or the Quantum Valley Hall effect (QVHE), have recently become an active field of investigation to overcome these limitations. While these systems seem promising from a practical point of view, the nature and role of topological protection is not well understood to date. Topological concepts such as the spin and valley Chern numbers for the QSHE and QVHE have entered the discussion. In electro-magnetic systems with reciprocity symmetry, the former is, however, neither rigorous invariant nor is a bulk-edge correspondence (in the Kane Mele sense \cite{PhysRevLett.95.226801}) established. The valley Chern number as defined in the literature \cite{chen2017valley,ma2016all}is not a topological invariant as shall become evident below, while a bulk-boundary correspondence based on a lattice-folding Hamiltonian has been recently established \cite{PhysRevB.98.155138}.

We here use a group theoretical pathway \cite{PhysRevLett.119.227401} to understand in which sense all-dielectric QSHE and QVHE edge modes are unidirectional and topologically protected. In particular, we show that QVHE edge modes in the center of the band gap are weakly protected against moderate arbitrary perturbations. In contrast to strongly $\mathbb{Z}$-protected integer quantum Hall states and $\mathbb{Z}_2$-protected proper quantum spin Hall states, their existence and robustness with respect to perturbations relies, however, on the particular choice of crystal termination and inclination. Backscattering immunity on the other hand requires orthogonality of counter-propagating \emph{chiral spin} states which is here based on spatial symmetries of the underlying bulk and shown to be approximately valid for small band gaps. We illustrate our findings by means of the QSHE PhC introduced in \cite{PhysRevLett.114.223901} and a new PhC design based on the kagome lattice \cite{Mekata2003,PhysRevB.80.113102}, and derive a list of general principles for the design of all-dielectric QVHE insulators for applications at optical or near-infrared frequencies.

\begin{figure}[t]
    \centering
    \includegraphics[width=1.\columnwidth]{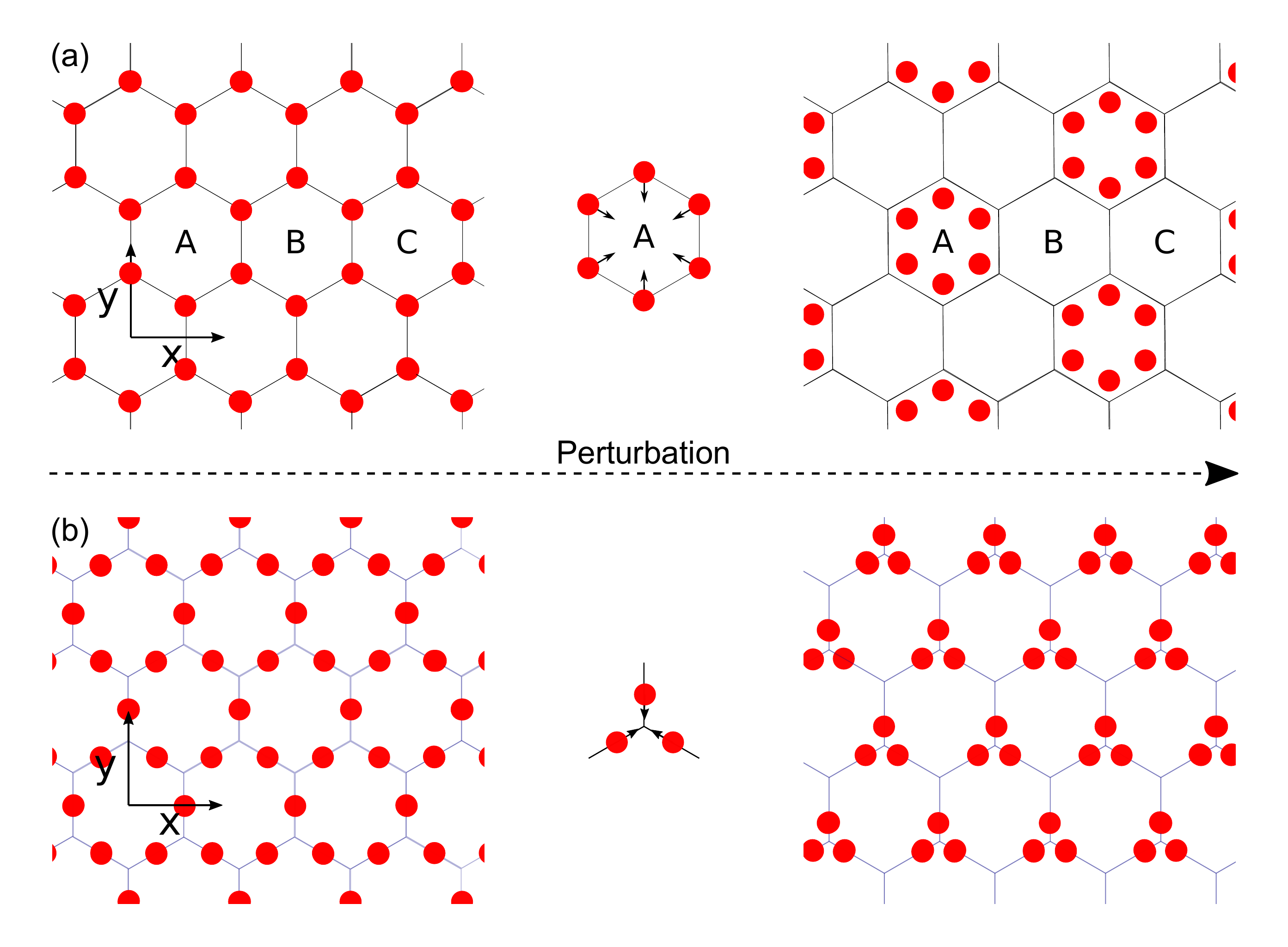}
    \caption{Two hexagonal lattice designs with broken $C_{6v}$ symmetry made of dielectric rods, indefinitely extended in the third direction. (a) QSHE honeycomb structure with sites at the corners of the hexagons. The design is perturbed by shrinking/expanding every third Wigner-Seitz cell labelled \textsf{A} \cite{PhysRevLett.114.223901}. (b) QVHE kagome structure with sites at the center of the edges. The design is perturbed by moving the rods along the edges.}
    \label{fig:honeycomb_kagome}
\end{figure}

The structure of this manuscript is as follows: In section \ref{sec:pert_C6v}, we explore the possibility of topological band gaps in symmetry perturbed $C_{6v}$ lattices. We group theoretically enumerate all possible spatial perturbations, excluding only long range chirps. A band gap opens only in three cases, one of which requires breaking of reciprocity symmetry, while the other two can be related to the well known QSHE and QVHE like systems (canonical examples of each case are shown in \figref{honeycomb_kagome} and discussed in detail in \cite{wong2019gapless}). The resulting bandstructures are derived and discussed in section \ref{sec:diag}. The topological properties of the QSHE and QVHE scenarios are illuminated in sections \ref{sec:QSHE} and \ref{sec:QVHE}, respectively. It is found that QSHE systems, while resembling a $\mathbb{Z}_2$ invariant edge state dispersion in first order perturbation, do not lead to topological protection. QVHE systems on the other hand lead to topologically protected edge modes that can be related to strong Chern protection. Since the type of protection only exists in an extended parameter space including geometrical perturbation, it depends on inclination and termination, consistent with the findings presented in \cite{PhysRevB.98.155138}. The associated bulk-boundary correspondence is therefore not comparable to strong protection through breaking of optical reciprocity \cite{PhysRevLett.100.013904}. The edge modes are, however, protected in the same way against moderate random perturbations if corresponding design principles are met.

\section{Perturbation theory on irreducible space group representations \label{sec:pert}}

{\renewcommand{\arraystretch}{1.5}
\begin{table*}
    \centering
    \begin{ruledtabular}
        \begin{tabular}{r|cccc}
            Term                     & Space Group Element $\hat{g}$                                                        & Hermiticity $\hat{O}_{\text{Herm}}$                  & Time Reversal $\hat{O}_\text{TR}$                & Reciprocity $\hat{O}_\text{Rec}$           \\ \hline
            $f_\lambda$~             & $ \cc{\Lambda}_{\vec{k}i}(g).f_\lambda.\Lambda_{\vec{k}i}^T(g)$                & $ f_\lambda^\dagger$         & $ \cc{\Lambda}_{\vec{k}i}(I).\cc{f_\lambda}.\Lambda_{\vec{k}i}^T(I)$
                                                                                                                      & $\cc{\Lambda}_{\vec{k}i}(I).f_\lambda^T.\Lambda_{\vec{k}i}^T(I)$ \\
            $\vec{f}_a$~             & $\Lambda_a^T(g)\cdot\left( \cc{\Lambda}_{\vec{k}i}(g).\vec{f}_a.\Lambda_{\vec{k}i}^T(g) \right)$
                                                                                                                      & $\vec{f}_a^\dagger$
                                                                                                                      & $\cc{\Lambda}_{\vec{k}i}(I).\cc{\vec{f}_a}.\Lambda_{\vec{k}i}^T(I)$
                                                                                                                      & $\cc{\Lambda}_{\vec{k}i}(I).\vec{f}_a^T.\Lambda_{\vec{k}i}^T(I)$ \\
            $\vec{f}_k$~            & $R(p)\cdot \left( \cc{\Lambda}_{\vec{k}i}(g).\vec{f}_k.\Lambda_{\vec{k}i}^T(g) \right)$
                                                                                                                      & $\vec{f}_k^\dagger$
                                                                                                                      & $- \cc{\Lambda}_{\vec{k}i}(I).\cc{\vec{f}_k}.\Lambda_{\vec{k}i}^T(I)$ 
                                                                                                                      & $- \cc{\Lambda}_{\vec{k}i}(I).\vec{f}_k^T.\Lambda_{\vec{k}i}^T(I)$ \\
        \end{tabular}
    \end{ruledtabular}
    \vspace{-.5em}
    \caption{Transformation of the different terms in \eqref{f_Taylor} under different symmetries of the system. Here, $f$ is understood as a second rank tensor in the double index $(s\alpha)$, and $(.)$ is the matrix product with respect to this double index; $\overline{f}$ denotes complex conjugation, $f^T$ the matrix transpose and $f^\dagger\eq(\overline{f})^T$ the conjugate transpose. $(\cdot)$ denotes the matrix product in the $\vec{a}$ and $\vec{k}$ vector space. $\Lambda_a(g)$ is the representation corresponding to the geometrical perturbation, and $R(g)$ the Rodrigues matrix representation for 3D Euclidean vectors. The parity or inversion symmetry $I$ covers the action of time reversal and reciprocity on the partners if $I$ is an element of the space group; if not, the action of these two symmetries maps out of the irreducible representation (irrep) in question and both do not yield any additional restriction on the form of $f$. Note that $I\eq C_2$ if a 2D planar group is considered instead of a 3D space group.}
    \label{tab:Transformations}
\end{table*}
}
We here derive a procedure to obtain the eigenvalues and fields close to a high symmetry point in the irreducible Brillouin zone (BZ), considering geometrical perturbations that break the original symmetry. The general idea is introduced in \cite{PhysRevLett.119.227401}. We here extend the procedure to space group symmetries (including translations) to deal with geometrical perturbations that break translation symmetry. This further requires simplification such that only group generators (rather than an infinite number of elements) need to be considered.

We start with a generic Hamiltonian $H(\vec{a},\lambda):\mathcal{H}\rightarrow\mathcal{H}$ that is a linear operator on a physical Hilbert space $\mathcal{H}$. An eigenvector $\vec{v}\,{\in}\,\mathcal{H}$ satisfies $H(\vec{a},\lambda)\,\vec{v}\,{=}\,0$, with the associated nonlinear eigenvalue $\lambda$ (the frequency for the monochromatic Maxwell equations) and free (geometrical) system parameters $a_i\,{\in}\, \mathbb{R}$. Let $H(\vec{a}{=}0,\lambda)$ be invariant under the action of space group symmetries $g\in\mathcal{G}$, \ie~$[g,H(0,\lambda)]\eq 0$ for all $g$ and $\lambda$, so that the eigenvectors are a superposition of partners $\varphi_{\vec{k}i;s\alpha}$ of a single irrep $\Lambda_{\vec{k}i}$ of the space group $\mathcal{G}$. For details on this statement and the group theoretical terminology used in the following see \cite{Bradley_GT,Dresselhaus,PhysRevLett.119.227401} and Appendix \ref{app:glossary}. In summary, $\vec{k}$ labels a star with respect to the invariant subgroup of translations $\mathcal{T}\,{\subset}\,\mathcal{G}$ and coincides with a Bloch wave vector within the irreducible BZ, \ie~the asymmetric unit cell of the associated reciprocal space; the index $i$ labels an irrep $\Delta_i(\vec{k})$ of the little group of $\vec{k}$, with partners labeled by $\alpha$. $\Lambda_{\vec{k}i}\eq\Delta_i(\vec{k})\uparrow\mathcal{G}$ is thus an induced representation with respect to the left-coset expansion $\mathcal{G} = \sum_s g_s \mathcal{T}\;(g_s\in\mathcal{G})$, with the partner index $s$ iterating over the inequivalent representations of $\mathcal{T}$ within the $\vec{k}$-star. These inequivalent representations are one-dimensional and have the Bloch character $\chi_s(T)\eq\exp\{\imath (p_s\,\vec{k})\cdot\vec{T}\}$ ($T{\in}\mathcal{T}$). $p$ is the isogonal point group element corresponding to $g$, \ie~in Seitz notation $g\eq\{p,t\}$, with $t$ a translation (not necessarily in $\mathcal{T}$ for non-symmorphic space groups).

The goal is to use perturbation theory in order to expand around an evaluation point $x=(\vec{a}\eq0,\lambda_0,\vec{k}_0)$. However, a perturbation approach can only be applied if the states close to the evaluation point are similar to the state at the evaluation point, or mathematically speaking that the Bures distance $D(\phi,\psi)=\sqrt{1-|\langle\phi|\psi\rangle|^2}$ (for properly normalized states $\langle \phi|\phi\rangle$ with the inner product $\langle\cdot|\cdot\rangle:\mathcal{H}^*\times\mathcal{H}\rightarrow \mathbb{C}$ associated with $\mathcal{H}$) approaches zero:
\begin{equation*}
    D\left(\psi(x),\,\psi(x+dx)\right) \overset{dx\rightarrow 0}{\rightarrow} 0\text{ .}
\end{equation*}
The orthogonality of the irrep's partners, on the other hand, implies that the distance of states (infinitesimally) apart in $\vec{k}$ is $1$, so that the above requirement is clearly violated. Separating a Bloch phase off the partners $\phi_{\vec{k} i;s\alpha} := u_{\vec{k} i;s\alpha}\,\exp\{\imath (p_s\,\vec{k})\cdot\vec{r}\}$ solves the dilemma if the new sesquilinear product is defined with $u$ instead of $\phi$. This, however, generally renders the Hamiltonian $\vec{k}$-dependent, with the mapped eigenproblem:
\begin{equation}
    \tilde{H}(\vec{a},\lambda,\vec{k})\, u_{\vec{k} i;s\alpha} \eq 0 \text{ .}
\end{equation}
To be precise, $\tilde{H}$ also depends on the partner index $s$. The following arguments do, however, not depend on this index, so that we drop it for the Hamiltonian in the vicinity of the evaluation point to avoid confusion with $s$ belonging to the irrep $\vec{k}_0 i$ at the evaluation point.

In zero order (degenerate) perturbation theory, the eigenvector at $\vec{k}_0+\delta\vec{k}$ in the vicinity of the evaluation point can be expressed as a linear combination $\sum_{s,\alpha} c_{s\alpha} u_{\vec{k}_0 i;s\alpha}$. Testing the eigenvalue equation at $\vec{k}_0+\delta\vec{k}$ with the partners $u_{\vec{k}_0i;s\alpha}$ thus yields a low-dimensional algebraic eigenvalue problem with the coefficients $c_{s\alpha}$ forming the eigenvectors:
\begin{equation}
    \sum_{s'\alpha'} \langle u_{s\alpha}\,|\,\tilde{H}(\delta\vec{k},\lambda_0+\delta\lambda,\vec{k}_0+\delta\vec{k})\, u_{s'\alpha'}\rangle c_{s'\alpha'} = 0 \text{ ,}
    \label{eq:eigen-c}
\end{equation}
where we have dropped the now redundant irrep index $(\vec{k}_0 i)$ and just kept the corresponding partner index $(s\alpha)$. The solution of \eqref{eigen-c} yields the bandstructure in the vicinity of the evaluation point, including the eigenmodes via $c_{s\alpha}$ and thus the topological charge of the evaluation point. In order to solve this equation without any explicit knowledge of the $u_{s\alpha}$, we first Taylor expand the matrix element 
\begin{align*}
    &f_{s\alpha}^{(s'\alpha')}(\vec{a},\lambda_0+\delta\lambda,\vec{k}_0+\delta\vec{k})\\
    &:= \langle u_{s\alpha}\,|\,\tilde{H}(\delta\vec{k},\lambda_0+\delta\lambda,\vec{k}_0+\delta\vec{k})\, u_{s'\alpha'}\rangle
\end{align*}
in first order:
\begin{align}
    f(\vec{a},\lambda_0+\delta\lambda,\vec{k}_0+\delta\vec{k}) &= f_0+\vec{f}_a\cdot\vec{a} + \vec{f}_k\cdot\delta\vec{k} + f_\lambda\,\delta\lambda \notag\\
    &=   \left[ \nabla_a f(\vec{a},\lambda_0,\vec{k}_0) \right]_{\vec{a}=0}\cdot\vec{a} \notag \\
    &\,+ \left[ \nabla_k f(0,\lambda_0,\vec{k}) \right]_{\vec{k}=\vec{k}_0} \cdot\delta\vec{k} \notag \\
    &\,+ \left[ \partial_\lambda f(0,\lambda,\vec{k}_0) \right]_{\lambda=\lambda_0} \delta\lambda \text{ .}
    \label{eq:f_Taylor}
\end{align}
In the last line we have used the fact that $f_0\eq0$ by definition.

We now employ our knowledge on how the different terms in \eqref{f_Taylor} transform under symmetry operations to derive \emph{selection rules} on the matrix elements. Consider a symmetry operation $\hat{O}$ that leaves the system Hamiltonian invariant, \ie~$[ \hat{O},H(\vec{a}\eq0,\lambda_0) ] \eq 0$. This implies that the matrix elements $f$ are also invariant under the action of $\hat{O}$. $f$ thus must be in the kernel
\begin{equation*}
    \left[ \hat{O} - \identity \right] f = 0 \text{ .}
\end{equation*}
Note that the symmetry operations act on $\delta\vec{k}$, but they leave $\vec{k}_0$ invariant by virtue of its definition as a label of the space group irreps (the action is instead implicitly included in the induced irreps themselves). For convenience, we wrap the action on the vectorial nature of the permutation parameters $\vec{a}$ and $\delta\vec{k}$ onto the matrix element in the scalar product: For example, we use the identity $\vec{f}_k\cdot(R^T(p)\delta\vec{k})\eq(R(p)\vec{f}_k)\cdot\delta\vec{k}$, so that the permutation parameter stays effectively invariant. This yields the independent kernel equations that need to be satisfied for all system symmetries $\hat{O}$:
\begin{align}
    \left[ \hat{O} - \identity \right] \vec{f}_a &= 0 \notag \\
    \left[ \hat{O} - \identity \right] \vec{f}_k &= 0 \notag \\
    \left[ \hat{O} - \identity \right] f_\lambda &= 0 \text{ .}
    \label{eq:gen_app}
\end{align}
Note that $f\,{\in}\,\text{GL}_n(\mathbb{C})$, so that $\vec{f}_k\,{\in}\, \mathbb{R}^d\,{\otimes}\,\text{GL}_n(\mathbb{C})$, where $d$ is the spatial dimension of the lattice (here $d\eq2$) and $n$ is the dimension of the space group irrep at the evaluation point. The system symmetry $\hat{O}$ thus acts on both vector spaces in the tensor product independently. The transformations under spatial and non-spatial symmetries are listed in \tabref{Transformations}. The non-spatial symmetries all square to the identity for spin-less systems. They are not independent and form a \emph{trinity}: Application of any pair yields the remaining operation, for example $\hat{O}_{\text{TR}}\hat{O}_{\text{Rec}}\eq\hat{O}_{\text{Herm}}$. For the unitary spatial symmetries $\hat{g}$, we show in \cite{thesis_Elman} that it suffices for the permutation elements to satisfy
\begin{align}
    \sum_{g\in\mathcal{S}}\left[ \hat{g} - \identity \right] \vec{f}_a &= 0 \notag \\
    \sum_{g\in\mathcal{S}}\left[ \hat{g} - \identity \right] \vec{f}_k &= 0 \notag \\
    \sum_{g\in\mathcal{S}}\left[ \hat{g} - \identity \right] f_\lambda &= 0 \text{ ,}
    \label{eq:gen_sum_app}
\end{align}
where $\mathcal{S}$ denotes an arbitrary set of generators of the space group $\mathcal{G}$. With this simplification, only a single kernel equation has to be satisfied for spatial symmetries per perturbation matrix element (instead of an infinite number of equations \eqref{gen_app}). Application of \eqref{gen_sum_app} for spatial symmetries and subsequently \eqref{gen_app} for non-spatial symmetries hence reduces the degrees of freedom in \eqref{f_Taylor} substantially and establishes its allowed form. After this step, we can solve \eqref{eigen-c} that takes the compact form:
\begin{equation}
    f(\vec{a},\delta\lambda,\delta\vec{k})\,\vec{c} = 0 \text{ .}
    \label{eq:pert_eig}
\end{equation}

\section{2D hexagonal lattices with $C_{6v}$ symmetry \label{sec:pert_C6v}}

The most systematic route towards topologically protected surface states in classical systems with lattice symmetry is breaking of symmetry induced (deterministic) degeneracies that carry a topological charge in (an extended) reciprocal space. A topologically non-trivial band gap can thus be opened. This idea has already been used in the original work by Haldane and Raghu where simultaneous breaking of reciprocity, time reversal symmetry and spatial symmetry has been suggested using a Faraday constituent material in a magnetic field \cite{PhysRevA.78.033834}. It has been exploited ever since in various systems \cite{Lu2014}.

In this work, we focus on systems with no broken reciprocity symmetry. There are two possible starting points for 2D wallpaper groups: First, accidental point-degeneracies with conical dispersion in square lattices with broken $C_{4v}$ symmetry, arising through splitting of flat 2-fold degeneracies at the $\Gamma$ and $X$ point in the original symmetry, are an intriguing starting point \cite{Makwana:19}. They are, however, more cumbersome to design, since the starting point is not a deterministic degeneracy (albeit stemming from one). We here instead focus on deterministic point-degeneracies with conical dispersion. These only exist at the $K$ ($K'$) point of hexagonal lattices, since all other deterministic degeneracies in 2D lattices are located at high symmetry points that are mapped onto themselves, immediately leading to flat bands if either reciprocity or time reversal symmetry are present.

{\renewcommand{\arraystretch}{1.5}
\begin{table}[t]
    \centering
    \begin{ruledtabular}
        \begin{tabular}{r|ccc}
            Representation         & $C_6$                     & $\sigma$                                & $T$ \\ \hline
            $\Lambda_{\Gamma1}$    & $1$                       & $1$                                     & $1$   \\
            $\Lambda_{\Gamma2}$    & $1$                       & $-1$                                    & $1$   \\
            $\Lambda_{\Gamma3}$    & $-1$                      & $1$                                     & $1$   \\
            $\Lambda_{\Gamma4}$    & $-1$                      & $-1$                                    & $1$   \\
            $\Lambda_{\Gamma5}$    & $R_6^2$                   & $\sigma_3$                              & $\identity_2$   \\
            $\Lambda_{\Gamma6}$    & $R_6$                     & $\sigma_3$                              & $\identity_2$   \\ \hline
            $\Lambda_{K1}$         & $\sigma_1$                & $\identity_2$                           & $\text{diag}\left( w^2,w \right)$ \\
            $\Lambda_{K2}$         & $\sigma_1$                & $-\identity_2$                          & $\text{diag}\left( w^2,w \right)$ \\
            $\Lambda_{K3}$         & $\sigma_1\otimes R_6^4$   & $\identity_2\otimes\sigma_3$            & $\text{diag}\left( w^2,w \right)\otimes\identity_2$ \\ \hline
            $\Lambda_{M1}$         & $P_{i+1}$                 & $\text{diag}\left( \sigma_1,1 \right)$  & $\text{diag}\left( -\identity_2,1 \right)$   \\
            $\Lambda_{M2}$         & $P_{i+1}$                 & $-\text{diag}\left( \sigma_1,1 \right)$ & $\text{diag}\left( -\identity_2,1 \right)$   \\
            $\Lambda_{M3}$         & $-P_{i+1}$                & $\text{diag}\left( \sigma_1,1 \right)$  & $\text{diag}\left( -\identity_2,1 \right)$   \\
            $\Lambda_{M4}$         & $-P_{i+1}$                & $-\text{diag}\left( \sigma_1,1 \right)$ & $\text{diag}\left( -\identity_2,1 \right)$
        \end{tabular}
    \end{ruledtabular}
    \vspace{-.5em}
    \caption{The irreducible space group representations in $p6mm$ that correspond to high symmetry points in the irreducible BZ. Listed are the representation matrices for the three group generators $C_6$, $\sigma$, and $T$ defined in the main text. For compactness of notation, $\sigma_i$ denotes the respective Pauli matrix, $R_6\defas1/2\left( \identity_2-\sqrt{3}\imath\sigma_y \right)$ the planar Rodrigues $60$ degrees rotation matrix, $w\defas\exp\{2\pi\imath/3\}$ the Bloch phase $p_{K'}(T)$, $P_{i+1}$ the 3D cyclic permutation matrix, $\identity_n$ the $n$-dimensional identity matrix, and $\text{diag}(\{e_i\})$ the block-diagonal matrix with the tuple of diagonal entries as argument.}
    \label{tab:2Dspacegroupirreps}
\end{table}}

We start with degeneracies at the $K$ point (note that the $K'$ point is in the same star and hence the same irrep of the wallpaper group) in a hexagonal wallpaper group $p6mm$ (17), which encompasses all other hexagonal wallpaper groups of lower symmetry. We will enumerate all (geometrical) perturbations $\vec{a}$ that break the $p6mm$ symmetry, and derive corresponding matrix elements $f$ (\eqref{f_Taylor}) that yield the dispersion and the bulk states in the vicinity of the $K$ point. To this end, we apply the formalism derived in Sec.~\ref{sec:pert} to establish the allowed form of the perturbation matrix $f(\vec{a},\delta\lambda,\delta\vec{k})$ in the vicinity of the evaluation point $(\vec{a},\lambda,\vec{k})\eq(0,\lambda_0,\vec{k}_0)$. 

\tabref{2Dspacegroupirreps} shows the irreps of $p6mm$ at the corners of the irreducible BZ. We here adopt the labelling of the little group irrep underlying the induced space group irreps from \cite{Bradley_GT}, \ie~in $\Gamma1$ correspond to $R_1$ of the little group $G^3_{12}$ in table 5.1 in \cite{Bradley_GT} \emph{etc}. At the $K$ point, the physical eigenfunctions that transforms according to $\Lambda_{K3}$ correspond to a 2-fold degeneracy in the bandstructure. As generators of the space group we choose the $C_6$ rotation, the mirror symmetry $\sigma$ that maps $y\mapsto-y$ and leaves $x$ invariant, and the translation $T$ along $x$, \ie~we choose the coordinate system shown in \figref{honeycomb_kagome}. As coset representatives $r_\alpha$ in the expansion $\mathcal{G}=\sum_\alpha r_\alpha \mathcal{G}_K$ (where $\mathcal{G}_K$ is the associated little group, \cf~also App.~\ref{app:glossary}), we choose the identity $E$ and the $C_2$ rotation.

For the sake of completeness, we first apply \eqref{gen_sum_app} to $\Lambda_{\Gamma5/6}$ which evidently describe 2-fold degeneracies at $\Gamma$. This procedure results in an empty nullspace and hence a vanishing $f_k$ matrix element in \eqref{f_Taylor}. In other words, the degeneracy does not split to first order in $\vec{k}$. This finding is a consequence of spatial symmetry alone, but we note that reciprocity does generally not allow linear splitting of a 2D degeneracy along all directions at $\Gamma$ (or any other point in the BZ that maps to itself under reciprocity). This can be seen from the fact that the first order $\vec{k}$-dependence of the 2D matrix $f$ can always be expressed as $f(\delta\vec{k})=\sum_{i=0}^3 \sigma_i \,P_i(\delta\vec{k})$, with $\sigma_0=\identity$ and the respective Pauli matrix for $i>0$, and $P_i$ a first order polynomial with vanishing constant term in the components of $\delta\vec{k}$, or $P_i=\vec{c}\cdot\delta\vec{k}$. Only $\sigma_2$ is thus commensurable with reciprocity (\cf~\tabref{Transformations}). The spectrum is hence given by $\delta\lambda\propto\text{eigs}\{\sigma_2P_2\}=\pm P_2$, \ie~it does not split at least along the line $P_2(\delta\vec{k})=0$ in $\vec{k}$ space.

Our main focus of interest is thus $\Lambda_{K3}$, for which application of \eqref{gen_sum_app} yields $f_k \propto \gamma_1\delta k_x - \gamma_2\delta k_y$, where we define a representation of the 4D Euclidean Clifford algebra as
\begin{align}
    \gamma_1 &:= \sigma_3\otimes\sigma_3 \notag\\
    \gamma_2 &:= \sigma_3\otimes\sigma_1 \notag\\
    \gamma_3 &:= \sigma_1\otimes\identity_2 \notag\\
    \gamma_4 &:= \sigma_2\otimes\identity_2 \notag\\
    \gamma_5 &:= \sigma_3\otimes\sigma_2 \text{ .}
    \label{eq:Clifford}
\end{align}
These matrices evidently satisfy the anticommutator relation $\{\gamma_i,\gamma_j\}\eq2\delta_{ij}$. Note that $f_k$ is invariant under non-spatial symmetries in \tabref{Transformations}, including time reversal symmetry and reciprocity, if the proportionality constant is chosen to be real. To see that we note that $\Lambda_{K3}(C_2)\eq\Lambda_{K3}^3(C_6)\eq \gamma_3$ according to \tabref{2Dspacegroupirreps}. Therefore, application of \tabref{Transformations} and the anticommutator relation for the Clifford matrices yields $\hat{O}_{\text{TR}}\gamma_{1/2}\eq\hat{O}_{\text{Rec}}\gamma_{1/2}\eq{-}\gamma_3.\gamma_{1/2}.\gamma_3\eq\gamma_{1/2}$.

\eqref{gen_sum_app} further trivially yields $f_\lambda\,{\propto}\,\identity$, while $f_a$ evidently depends on the type of geometrical perturbation. We first discuss the two cases of interest that lead to the QSHE and QVHE effect, respectively, and then show that those are indeed the only non-trivial cases among all geometrical perturbations without long-range chirp \cite{Hielscher:17}, \ie~those that correspond to the high symmetry points in the irreducible BZ listed in \tabref{2Dspacegroupirreps} with $\Lambda^n(T)\eq\identity$ for some finite integer $n$.

\begin{figure}[t]
    \centering
    \includegraphics[width=\columnwidth]{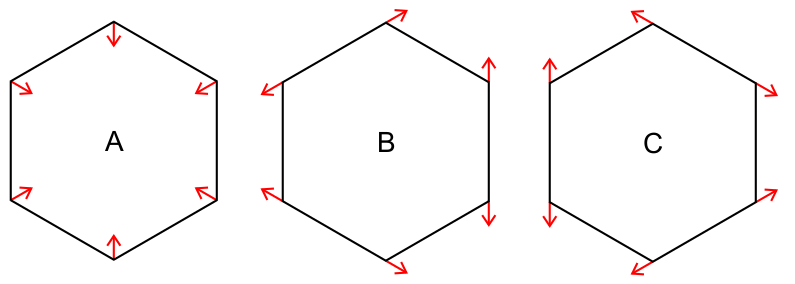}
    \caption{The perturbation arrows of the six closest lattice sites from the origin for the three different partners A,B,C belonging to the QSHE perturbation. The honeycomb hexagon is shown as a guide to the eye. The whole lattice is spanned by application of the reduced translation symmetry (\cf~\figref{honeycomb_kagome}).}
    \label{fig:QSHE_partners}
\end{figure}

We start with the perturbation introduced in \cite{PhysRevLett.114.223901} and illustrated in \figref{honeycomb_kagome} (a). The representation $\Lambda_a$ corresponding to this perturbation is most conveniently chosen to be three-dimensional, with partners illustrated in \figref{QSHE_partners}. By inspection, the representation matrices for the group generators are then obtained as
\begin{align*}
    \Lambda_{\text{QSHE}}(C_6) &= 
    \begin{pmatrix}
        1 & 0 & 0 \\
        0 & 0 & 1 \\
        0 & 1 & 0
    \end{pmatrix}\text{,}\\
    \Lambda_{\text{QSHE}}(\sigma) &= \identity\text{, and}\\
    \Lambda_{\text{QSHE}}(T) &= 
    \begin{pmatrix}
        0 & 1 & 0 \\
        0 & 0 & 1 \\
        1 & 0 & 0
    \end{pmatrix} \text{.}
\end{align*}
Application of \eqref{gen_sum_app} for the spatial symmetries leads to
\begin{align*}
    f_a &= d_1 \identity_4 \left( a_A+a_B+a_C \right) \\ &+
    d_2 \Big[ \begin{pmatrix}
        0_{2x2} & \identity_2 \\
        \identity_2 & 0_{2x2}
    \end{pmatrix} a_A \\ &+
    \begin{pmatrix}
        0_{2x2} & w\identity_2 \\
        w^2\identity_2 & 0_{2x2}
    \end{pmatrix} a_B \\ &+
    \begin{pmatrix}
        0_{2x2} & w^2\identity_2 \\
        w\identity_2 & 0_{2x2}
    \end{pmatrix} a_C \Big]\text{ ,}
\end{align*}
with $w\defas\exp\{2\pi\imath/3\}$, $\identity_n$ the $n{\times}n$-dimensional identity matrix, and $0_{n{\times}m}$ the $n{\times}m$-dimensional zero matrix. This result is commensurable with the non-spatial symmetries if $d_1,d_2\,{\in}\,\mathbb{R}$ by application of \tabref{Transformations} and the fact that $\vec{a}\,{\in}\,\mathbb{R}^3$ by definition. We disregard the trivial term that shifts all bands simultaneously up or down without altering eigenvectors in the following and set $d_1=0$. Additionally, $a_B$ and $a_C$ have only been introduced to obtain a full set of partners under the symmetry operations of the unperturbed lattice, while the deterministic perturbation in \cite{PhysRevLett.114.223901} only uses $a_A\asdef a$. We will thus in the following further set $a_B\eq a_C\eq 0$, finally leading to the simplified form
\begin{equation}
    f_{\text{QSHE}}\left(a,\lambda, \delta\vec{k} \right) = \gamma_1\delta k_x - \gamma_2 \delta k_y + \gamma_3 a - \identity \lambda \text{ ,}
    \label{eq:QSHE_perturbationHamiltonian}
\end{equation}
with the perturbation parameters rescaled such that the respective coefficients are absorbed and \eqref{QSHE_perturbationHamiltonian} is in the most convenient form. A different choice would not alter any of the findings below, and only lead to a different global slope of the bands and a different sign of the topological invariants which globally depends on the particular physical realisation and the choice of origin and direction of the arrows in \figref{QSHE_partners}. We will specifically discuss the dependence on the choice of origin below which clearly distinguishes the topological protection found here from strong protection in the integer quantum Hall sense.

{\renewcommand{\arraystretch}{1.5}
\begin{table}[t]
    \centering
    \begin{ruledtabular}
        \begin{tabular}{r|ccc}
            Irrep                  & spatial only                                               & including non-spatial                                     & gap  \\ \hline
            $\Lambda_{\Gamma1}$    & $\identity_4$                                              & $\identity_4$                                             & no        \\
            $\Lambda_{\Gamma2}$    & $\identity_2\otimes\sigma_2$                               & $-$                                                       & (yes)     \\
            $\Lambda_{\Gamma3}$    & $\sigma_3\otimes\identity_2$                               & $-$                                                       & no        \\
            $\Lambda_{\Gamma4}$    & $\gamma_5$                                                 & $\gamma_5$                                                & yes       \\
            $\Lambda_{\Gamma5}$    & $(\identity_2\otimes\sigma_3,\identity_2\otimes\sigma_1)$  & $(\identity_2\otimes\sigma_3,\identity_2\otimes\sigma_1)$ & no        \\
            $\Lambda_{\Gamma6}$    & $(\gamma_1,\gamma_2)$                                      & $-$                                                       & no        \\ \hline
            $\Lambda_{K1}$         & $\gamma_3\mp\imath\gamma_4$                                & $\gamma_3$                                                & yes       \\
            $\Lambda_{K2}$         & $(\sigma_1\mp\imath\sigma_2)\otimes\sigma_2$               & $-$                                                       & no        \\
            $\Lambda_{K3}$         & $\big((\sigma_1-\imath\sigma_2)\otimes\sigma_3,$               & $\big(\sigma_1\otimes\sigma_3,$                       &           \\
                                   & $\phantom{\big(}(\sigma_1-\imath\sigma_2)\otimes\sigma_1,$     & $\phantom{\big(}\sigma_1\otimes\sigma_1,$             & no        \\    
                                   & $\phantom{\big(}(\sigma_1+\imath\sigma_2)\otimes\sigma_3,$     & $\phantom{\big(}\sigma_2\otimes\sigma_3,$             &           \\
                                   & $\phantom{\big(}(\sigma_1+\imath\sigma_2)\otimes\sigma_1\big)$ & $\phantom{\big(}\sigma_2\otimes\sigma_1\big)$         &          \\ \hline
            $\Lambda_{M1}$         & $-$                                                        & $-$                                                       & $-$         \\
            $\Lambda_{M2}$         & $-$                                                        & $-$                                                       & $-$         \\
            $\Lambda_{M3}$         & $-$                                                        & $-$                                                       & $-$         \\
            $\Lambda_{M4}$         & $-$                                                        & $-$                                                       & $-$
        \end{tabular}
    \end{ruledtabular}
    \vspace{-.5em}
    \caption{List of the terms $f_a$ corresponding to geometrical perturbations that transform according to irreducible planar group representations in \tabref{2Dspacegroupirreps}. The first column lists the tuple of matrices corresponding to the partners defined through the representation matrices in \tabref{2Dspacegroupirreps} for spatial symmetries only. If non-spatial symmetries (time reversal and reciprocity) are included, only a specific linear combination of partners might be admissible, thus leading to a reduced tuple size (for example for $\Lambda_{K1}$, where only a perturbation along $a_1\,{+}\,a_2$ leads to non-zero $f_a\,{\propto}\,\gamma_3$). The coefficients of the tuples including non-spatial symmetries further are assumed to be real numbers, and ``$-$'' indicates that no perturbation matrix is found to satisfy all symmetries. The last column lists the possibility of a topological bandgap under the particular perturbation.}
    \label{tab:generic_pert_matrices}
\end{table}}

Substitution of \eqref{QSHE_perturbationHamiltonian} into \eqref{pert_eig} yields the $4$-dimensional linear eigenvalue problem
\begin{equation}
    W_{\text{QSHE}}\, \vec{c} = \left( \gamma_1\delta k_x - \gamma_2\delta k_y + \gamma_3 a \right) \vec{c} = \lambda\, \vec{c} \text{ .}
    \label{eq:QSHE_eigen}
\end{equation}
We will diagonalize the \emph{perturbation Hamiltonian} $W$ in the following section. We here proceed with the QVHE case. Consider the perturbed kagome lattice \cite{Mekata2003} illustrated in \figref{honeycomb_kagome} (b). The representation matrices belonging to the perturbation are evidently one-dimensional and given by
\begin{align*}
    \Lambda_{\text{QVHE}}(C_6) &= -1\text{ ,}\\
    \Lambda_{\text{QVHE}}(\sigma) &= -1\text{ , and}\\
    \Lambda_{\text{QVHE}}(T) &=  1\text{ .}
\end{align*}

Following the same procedure as above, the corresponding perturbation eigenproblem is shown to be
\begin{equation}
    W_{\text{QVHE}}\, \vec{c} = \left( \gamma_1\delta k_x - \gamma_2\delta k_y + \gamma_5 a \right) \vec{c} = \lambda\, \vec{c} \text{ .}
    \label{eq:QVHE_eigen}
\end{equation}
Let us now revisit the nature of the two particular example systems that lead to the QSHE and QVHE perturbation Hamiltonians \eqref{QSHE_eigen} and \eqref{QVHE_eigen} above. The question we are asking here in particular is whether there are other possible perturbations leading to a different Hamiltonian, respectively. Any arbitrary perturbation can be understood as belonging to some vector space that is spanned by a choice of basis vectors. These can be understood as partners of a generally reducible representation of $\mathcal{G}$, and thus can be decomposed into irreps. The completeness of irreps \cite{Dresselhaus,PhysRevB.88.245116} therefore guarantees that any perturbation transforms according to a direct sum of irreps of $\mathcal{G}$, which reduces to the irreps listed in \tabref{2Dspacegroupirreps} since we exclude long-range chirps that would lead a system with no translational symmetry.

For example, we can immediately identify $\Lambda_{\text{QVHE}}=\Lambda_{\Gamma4}$. For $\Lambda_{\text{QSHE}}$, we introduce the similarity transform $S^{-1}.\Lambda_{\text{QSHE}}(g).S$, with
\begin{equation*}
    S := 
    \begin{pmatrix}
        1&1&1\\1&w&w^2\\1&w^2&w
    \end{pmatrix}\text{ .}
\end{equation*}
This simultaneously brings the generator matrices (and thus all representation matrices) into an irreducible block form. Comparison of the blocks with \tabref{2Dspacegroupirreps} immediately reveals the direct sum partition $\Lambda_{\text{QSHE}}=\Lambda_{\Gamma1}\oplus\Lambda_{K1}$. We can now interpret our previous result for the QSHE $f_a$ in the following way: The trivial identity matrix part originates from the trivial planar group representation $\Lambda_{\Gamma1}$ (the same that lead to $f_\lambda\,{\propto}\,\identity$), while the non-trivial part stems from $\Lambda_{K1}$.

Using the above conclusion that any geometrical perturbation of interest is a direct sum of the planar group representations listed in \tabref{2Dspacegroupirreps}, we can now build the most general perturbation matrix by considering these irreps separately. Note that the reverse is not necessarily true, \ie~not all direct sums of the irreducible representations in \tabref{2Dspacegroupirreps} can be interpreted as a geometrical perturbation, as the latter requires $\Lambda_a(g)\in\text{GL}_n(\mathbb{R})$. For example, the $\Lambda_{Ki}$ on their own cannot form a geometrical perturbation as the generator matrix $\Lambda_{Ki}(T)\,{\notin}\,\text{GL}_n(\mathbb{R})$ (and there is no similarity transform that makes all generator matrices real).

\section{Diagonalisation of perturbation Hamiltonians \label{sec:diag}}

Consider the general perturbation Hamiltonian
\begin{equation}
    W = \gamma_1 \delta k_x - \gamma_2 \delta k_y +f_a a
    \label{eq:generic_W}
\end{equation}
as derived in section \ref{sec:pert_C6v}. Generally, the eigenvalues $\lambda_\alpha$ corresponding to $W$ are solutions of the characteristic equation $P(\lambda,\delta k_x,\delta k_y,a)=0$, where $P$ is a quartic polynomial. Since the main goal is to obtain frequency isolated topological edge states, a necessary requirement on $f_a$ is that it produces a gapped dispersion relation, for which $\lambda_\alpha(\delta k_x,\delta k_y,a)\,{\ne}\,0$ for all $\alpha$ and $\delta\vec{k}$ if $a\,{\ne}\,0$.
\begin{figure}[t]
    \centering
    \includegraphics[width=\columnwidth]{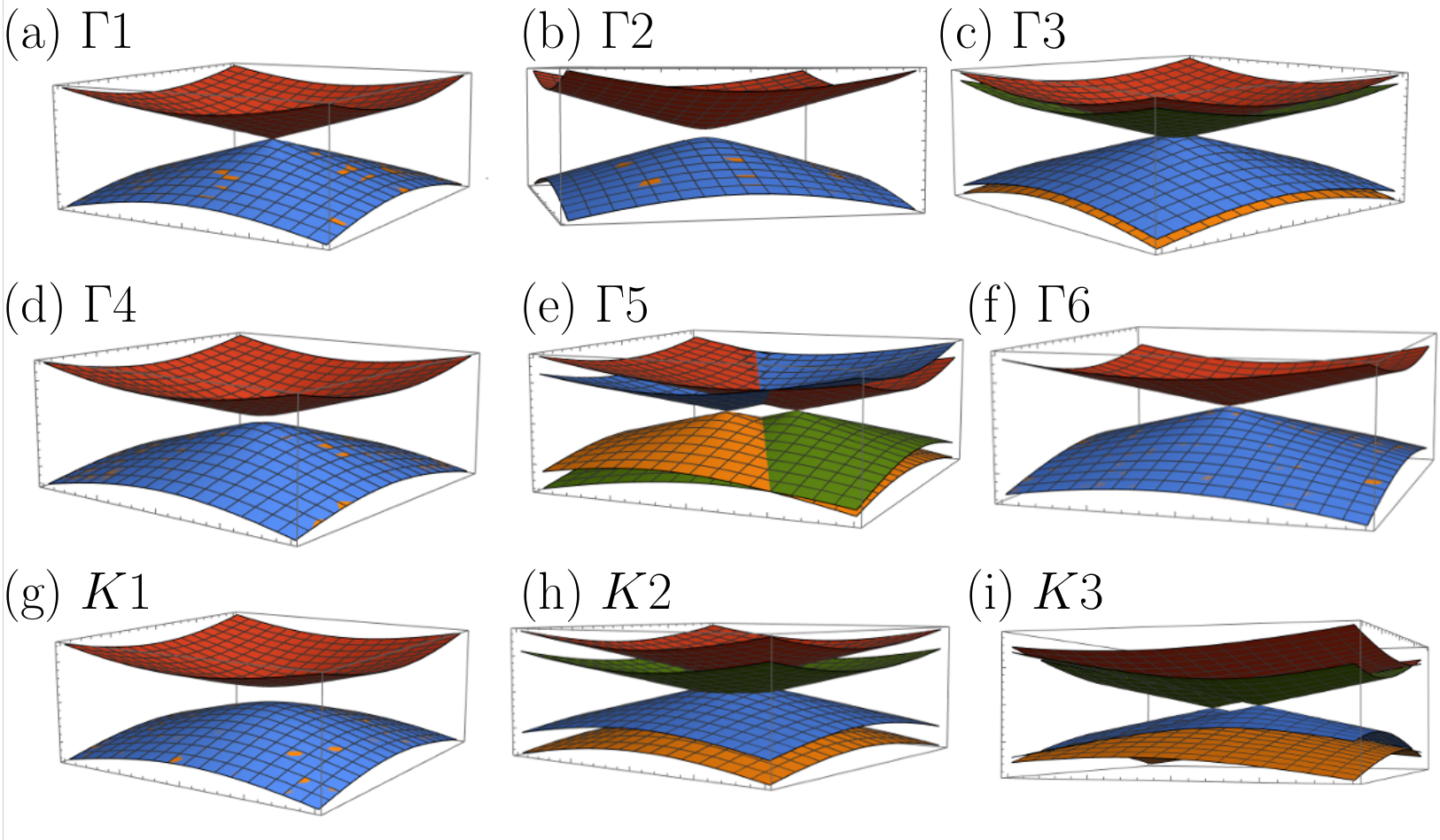}
    \caption{Geometrical perturbation bandstructures in the vicinity of the $K$ space group irrep. (a)-(f) perturbations that are trivial under translations, (g)-(i) perturbations that act similar to $K$ point irreps as involved in the QSHE perturbation Hamiltonian in \eqref{QSHE_eigen} (\cf also \tabref{generic_pert_matrices}). The bandstructures shown are for the cases without time-reversal symmetry and reciprocity enforced, and arbitrary perturbation. A band gap for finite perturbation only opens up in case (b), (d), and (g), and only for the latter two cases in systems with time-reversal symmetry.}
    \label{fig:perturbation_BS}
\end{figure}

Generally, any $4{\times}4$ matrix $M$ over the complex numbers can be represented by $M\eq\sum_{n,m=0}^3 \gamma_{(nm)}\,c_{(nm)}$, with $\gamma_{(nm)}\defas\sigma_n\otimes\sigma_m$, $\sigma_0\defas\identity_2$, and $c_{(nm)}\in\mathbb{C}$. This can be easily seen by considering $M_{(\alpha\beta)}$ and $c_{(nm)}$ as $16$-dimsensional vectors, respectively. The $16\times16$ matrix $\gamma_{(\alpha\beta),(nm)}$ above has evidently full rank (the four $\sigma$ matrices are linearly independent) so that the $c_{(nm)}$ can be explicitly constructed by inversion of $\gamma$. However, the derived perturbation Hamiltonian matrices summarized in \tabref{generic_pert_matrices} only contain a single of those $\gamma$ matrices per perturbation partner. This substantially reduces the complexity of the eigenproblem. In the following, we use that two $\gamma$ matrices either commute or anti-commute, and the fact that if $\lambda\in\text{eig}(W)$ we also have $\lambda^2\in\text{eig}(W^2)$. Let us now study the Hamiltonian $W\eq\gamma x+\gamma' y$, and discuss the two cases I: $[\gamma,\gamma']\eq0$ and II: $\{\gamma,\gamma'\}\eq0$. Case I:
\begin{align*}
    W^2 &= \left( x^2+y^2 \right)\identity + 2 xy\gamma\gamma' \\
    \Rightarrow \lambda^2 &= x^2+y^2+2\chi xy \text{ ,}
\end{align*}
where $\chi\in\text{eig}(\gamma\gamma')$ is a fourth root of $1$. If $W$ is hermitian, we immediately see that $\chi=\pm1$, so that
\begin{equation*}
    \lambda = \pm\sqrt{x^2+y^2\pm2xy}=\pm|x\pm y|\text{ .}
\end{equation*}
Since $W$ is also chirally symmetric with respect to some $\gamma''$ that anti-commutes with $\gamma$ and $\gamma'$ (we can always find $5$ matrices satisfying a Clifford algebra), \ie~$\gamma''\gamma\gamma''=-\gamma$ and equivalent for $\gamma'$, we further know that the eigenvalues come in pairs $\text{eig}(W)\eq\{\pm\lambda_1,\pm\lambda_2\}$. In other words, the eigenvalues above split symmetrically into two positive and two negative branches that meet along the lines $x\eq\pm y$ in the 2D parameter space, respectively, with eigenvalue $\lambda_{1/2}=0$. We thus immediately see that a combination of commuting $\gamma$ matrices does not allow for the desired complete bandgap in the vicinity of the evaluation point.
Case II yields the much simpler result
\begin{align*}
    W^2 &= \left( x^2+y^2 \right)\identity \\
    \Rightarrow \lambda^2 &= x^2+y^2 \text{ ,}
\end{align*}
which corresponds to a doubly-degenerate and conical dispersion in the vicinity of the evaluation point. To avoid confusion, we remind the reader that we are working on the basis of space group representations with index $\vec{k}$, which is closely related to, but not identical to the Bloch wave vector in the first BZ. Thus, if a doubly-degenerate hyperconic dispersion is described here without breaking translation symmetry (as in $\Lambda_{\Gamma n}$), this translates to two single hypercones at $K$ and $K'$ in the standard Bloch function picture, respectively.

The same analysis can be performed for a three or four term Hamiltonian, as appearing in \eqref{generic_W}, leading to typical bandstructures for finite fixed perturbations as shown in \figref{perturbation_BS}. Noticeably, only in three cases (only two with time-reversal symmetry) a bandgap opens for all $\vec{k}$ close to the evaluation point: Both of the time-reversal/reciprocity symmetric scenarios correspond to the previously discussed QVHE ($\Gamma4$) and QSHE ($K1$) cases, while the third case ($\Gamma2$) is the well-known reciprocity-breaking scenario \cite{PhysRevLett.100.013904}. We have thus rigorously shown that there are only the two well-known systems that have a chance to lead to frequency isolated topological edge modes in hexagonal lattices (as long as long-range compressions of the whole structure that destroy periodicity are excluded). While this might not come as a surprise given the amount of published work on the topic, our analysis clearly shows that looking for alternative routes based on geometrically perturbed hexagonal symmetries is an endeavour doomed to failure.

\section{Ostensible $\mathbb{Z}_2$ topological charge in QSHE systems \label{sec:QSHE}}

In this section, we study the QSHE Hamiltonian in-depth. In this case, the original translation symmetry of the unperturbed $p6mm$ lattice is broken. The interpretation of the partners is thus that these correspond to the trivial representation with respect to the translation group of the perturbed structure, \ie $\delta\vec{k}$ is with respect to the $\Gamma$ point for all partners in the standard bandstructure picture of the perturbed symmetries BZ. The bulk dispersion for finite perturbation strength $a$ is reminiscent of that in \figref{perturbation_BS} (g). In order to understand the occurrence of surface bands and the associated topology, we need to examine the eigenvectors $\vec{c}$ of $W_{\text{QSHE}}$ that solve $\eqref{QSHE_eigen}$.

In the following, we make use of the fact that $W_{\text{QSHE}}$ evidently commutes with $S\eq\sigma_1\otimes\sigma_2$ for any $\left( \delta k_x,\delta k_y,a \right)$, and that the eigenstates can now be obtained from the eigenfunctions of $S$. In Fermionic systems $S$ has the meaning of time reversal invariance, corresponding to an anti-unitary operator with $S^2\eq {-}1$ and associated Kramers pairs. We here have a simple linear operator that plays a similar role that is, however, understood entirely within the framework of linear maps. The spin channels are spanned by the two separate eigenvalues of geometric multiplicity of two of $S$, respectively. In the following, we need to bear in mind, however, that $S$ only commutes with the first order perturbation Hamiltonian, and the associated Spin Chern number is thus no genuine topological invariant of the system.

The eigensolutions of $S$ are (where $n$ accounts for geometric multiplicity and we denote the two induced orthogonal channels by $\uparrow$, $\downarrow$ respectively):
\begin{equation*}
    [\lambda_S;\vec{v}_n] = 
    \begin{cases}
        \left[ -1;\,
            \frac{1}{\sqrt{2}} \begin{pmatrix}  \imath\\0\\0\\1 \end{pmatrix},
            \frac{1}{\sqrt{2}} \begin{pmatrix} 0\\-\imath\\1\\0 \end{pmatrix}
        \right],& \uparrow\\
        \left[ \phantom{-}1;\,
            \frac{1}{\sqrt{2}} \begin{pmatrix} -\imath\\0\\0\\1 \end{pmatrix},
            \frac{1}{\sqrt{2}} \begin{pmatrix} 0\\ \imath\\1\\0 \end{pmatrix}
        \right],& \downarrow
    \end{cases}\text{ .}
\end{equation*}
The two different eigenvalues thus indeed induce two orthogonal 2D subspaces. Due to $[S,W_{\text{QSHE}}]\eq0$, there hence exists a linear combination of the two states in each channel that solves the original eigenproblem \eqref{QSHE_eigen}, \ie the associated kernel of the overdetermined problem
\begin{align*}
    \text{ker}\left[(W_{\text{QSHE}}-\lambda_{\pm}\identity_4).(\vec{v}_1,\vec{v}_2)\right] =
    \begin{cases}
        \begin{pmatrix}
            \lambda_{\pm}+\delta k_x\\ \delta k_y+\imath a
        \end{pmatrix},& \uparrow\\
        \begin{pmatrix}
            \lambda_{\pm}+\delta k_x\\ \delta k_y-\imath a
        \end{pmatrix},& \downarrow
    \end{cases}
\end{align*}
is not empty. In the above, $\lambda_{\pm}\,{=}\,\pm\sqrt{\delta k_x^2+\delta k_y^2+a^2}$ are the established eigenvalues of the original eigenproblem. Comparing with \eqref{pert_eig}, the calculation of the Berry connection or curvature of the physical Hilbert space collapses to the 4D space spanned by the irreducible partners \cite{PhysRevLett.119.227401}, and here to the 2-dimensional vector space induced by $S$ since the spin basis vectors are $\vec{k}$-independent and ortho-normalized. The 2D vectors above are, on the other hand, the (non-normalized) eigenvectors of a Weyl Hamiltonian of opposite chirality and Chern integer $C_{\uparrow\pm}\eq\mp 1$ and $C_{\downarrow\pm}=\pm1$.

A $\mathbb{Z}_2$ Kane Mele bulk-boundary correspondence would thus be guaranteed by the fact that the Berry curvature vector is symmetric with respect to the mirror operation $a\,{\mapsto}\, {-}a$, so that the spin Chern number is $C\eq C_{\uparrow\pm}-C_{\downarrow\pm}\eq{\mp} 1$ when integrated over the finite perturbation plane $a=\text{const.}>0$.\footnote{We should of course integrate over the finite hexagonal BZ, but the argument does not change by this slight complication since the total Berry curvature of all Weyl points through the $a=a_0$ plane in an extended zone scheme is $\pm N/2$ for $N$ BZs, so that the integrated Berry curvature over one BZ evaluates to $\pm1/2$ after taking the limit $N\rightarrow \infty$ and using translation symmetry.} As stated above, however, the whole argument is based on the vector-space spanned by the irreps of the evaluation point and thus valid in zero order perturbation theory only. The problem can be easily seen when considering the bandstructure of a mixed QSHE-QVHE Hamiltonian $W\eq\gamma_1 k_x-\gamma_2k_y+\gamma_3a+\gamma_5b$ within the 2D parameter space $(a,b)\in\mathbb{R}^2$. It is well known that the (spin) Chern number's phase boundaries coincide with a closing of the associated band gap \cite{PhysRevA.78.033834}. If the spin Chern number is a proper topological invariant of the bulk, we can thus assign a phase space function $C(a,b){:}\, \mathbb{R}^2\,{\rightarrow}\,\{-1,1\}$. The bandgap, however, only closes at $(a,b)\eq0$ because all $4$ terms in $W$ are part of the Clifford algebra \eqref{Clifford}. Hence, $C(a,b)$ is constant in $\mathbb{R}^2\setminus 0$. But we have computed $C\eq{\pm}1$ for $a\,{\lessgtr}\,0$ on the $b\eq0$ axis above. The paradox is trivially resolved by realizing that the spin Chern number $C$ above is only an integer for $b\eq0$ in the first place (since $[S,\gamma_5]\ne 0$). We have thus shown that the spin Chern number is generally no longer a topological invariant if the system is perturbed in any other way additionally to the QSHE direction (even when staying within first order perturbation theory).

We nevertheless expect surface states following a QSHE dispersion in the vicinity of the $\Gamma$ point if the perturbation strength $a$ is small, \ie~for small band gaps, with counter-propagating modes that are approximately orthogonal and hence non-interacting. To elaborate on this statement, we consider the surface of two crystals with perturbation $\pm a$ in the positive/negative half-space in $x$. Note that we did not assume the parameters to be real numbers during the derivation above. They hold for imaginary parameters, and in particular for half-space solutions obtained by Floquet theory, \ie~a transfer matrix approach. Let us thus assume a complex wave number in the positive [negative] half-space such that the associated fields are normalizable, with $\Im\{\delta k_x\}\geq\left[ \leq \right]0$. A minimum requirement for the existence of a surface state is that the half-space solutions match at the interface $x{=}0$. In the upper spin channel they are proportional to (in the basis of upper spin vectors, and with $\delta k_x$ the wave number in the positive half-space)
\begin{equation*}
    \begin{pmatrix}
        \lambda + \delta k_x \\ \delta k_y + \imath a
    \end{pmatrix}\quad\text{and}\qquad
    \begin{pmatrix}
        \lambda - \delta k_x \\ \delta k_y - \imath a
    \end{pmatrix}
\end{equation*}
in the positive and negative half-space, respectively. Coupling thus yields $\lambda\eq\delta k_y$ and $\delta k_x\eq\imath a$.
For the lower spin channel we have
\begin{equation*}
    \begin{pmatrix}
        \lambda + \delta k_x \\ \delta k_y - \imath a
    \end{pmatrix}\quad\text{and}\qquad
    \begin{pmatrix}
        \lambda - \delta k_x \\ \delta k_y + \imath a
    \end{pmatrix}
\end{equation*}
and obtain $\lambda\eq-\delta k_y$ instead. The two associated normalized eigenvectors are
\begin{equation*}
    \frac{1}{\sqrt{2}}\begin{cases}
\begin{pmatrix}
    1\\1
\end{pmatrix},&\uparrow\\
\begin{pmatrix}
    1\\-1
\end{pmatrix},&\downarrow
    \end{cases}\text{ .}
\end{equation*}
Two orthogonal surface states with linear surface band dispersion of opposite slope, and a decay length that is inversely proportional to the perturbation strength, are thus predicted. The same result applies to any orthogonal pair of $(k_\parallel,k_\perp)$ in the reciprocal plane, easily seen by substituting $\delta k_x\eq k_\perp \cos\varphi-k_\parallel\sin\varphi$ and $\delta k_y\eq k_\perp\sin\varphi+k_\parallel\cos\varphi$ for some $\varphi\in[0,2\pi)$.

\section{Genuine $\mathbb{Z}$ charge in QVHE systems \label{sec:QVHE}}

The QVHE Hamiltonian
\begin{equation*}
    W_{\text{QVHE}} = \gamma_1 \delta k_x - \gamma_2 \delta k_y + \gamma_5 a
\end{equation*}
is algebraically similar to the QSHE Hamiltonian. It thus leads to the same hyperconic dispersion, and similar spin separation. The spin separation, however, is mathematically trivial here, as $W_{\text{QVHE}}=\sigma_3\otimes W_2$ naturally separates through a Kronecker product so that the outer vector space can be associated with the two spin channels. In other words, the generating spin operator might be chosen as $S=\sigma_3\otimes \identity$. The eigenpairs are:
\begin{equation*}
    [\lambda_S;\vec{v}_n] = 
    \begin{cases}
        \left[ -1;\,
            \begin{pmatrix}  0\\0\\1\\0 \end{pmatrix},
            \begin{pmatrix}  0\\0\\0\\1 \end{pmatrix}
        \right],& \downarrow\\
        \left[ \phantom{-}1;\,
            \begin{pmatrix} 1\\0\\0\\0 \end{pmatrix},
            \begin{pmatrix} 0\\1\\0\\0 \end{pmatrix}
        \right],& \uparrow
    \end{cases}\text{ .}
\end{equation*}
The derivation of surface state dispersion close to the $K(K')$ point is formally equivalent to the last section. The crucial difference lies in the interpretation of this finding: Translation symmetry is not broken in the QVHE scenario, so that, going back to the traditional bandstructure picture, the two spin channels correspond to the $K$ and $K'$ points in the BZ, respectively.

Importantly, we can now look at the question of topological protection from a different perspective. Before proceeding with this idea, let us formalise the concept of topological Chern (or $\mathbb{Z}$) protected edge states. We here consider hermitian Hamiltonians only.\vspace{1em}\\
\textbf{Bulk invariant: }Consider a physical eigenproblem with $d$-dimensional lattice symmetry, with an associated family of (normalized) eigenvectors $v(\vec{k})$, where $\vec{k}$ is an element of the reciprocal d-torus ($\vec{k}$-space) induced by the lattice. The family of eigenvectors thus forms a smooth vector bundle over $\vec{k}$-space. Any closed 2D manifold $\mathcal{M}$ within $\vec{k}$-space (with tangent space parametrization $(k_1,k_2)$) thus satisfies the generalized Gauss-Bonnet theorem \cite{chern1945curvatura}
\begin{equation*}
    \int_M \Omega = 2\pi\, \mathcal{C}
\end{equation*}
for the associated Berry curvature
\begin{equation*}
    \quad \Omega = -2\, dk_1\wedge dk_2\;\Im\langle \partial_{k_1} v, \partial_{k_2}v \rangle\text{ ,}
\end{equation*}
\begin{figure}[t]
    \centering
    \includegraphics[width=\columnwidth]{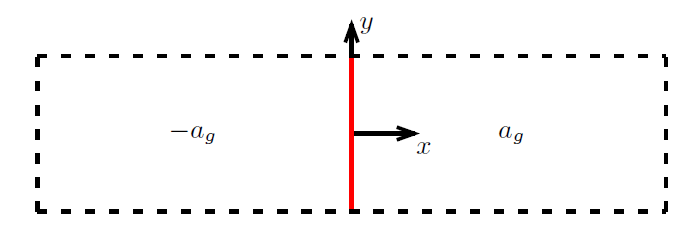}
    \vspace{-2em}
    \caption{Setup for QVHE strong protection. Two Semi-infinite half-space domains are perturbed with opposite perturbation strength $a_g$, separated by a (red) domain wall pointing in $y$ direction.}
    \label{fig:ky}
\end{figure}
with the Chern (aka Euler) number $\mathcal{C}\,{\in}\,\mathbb{Z}$ that is a topological integer.\vspace{1em}\\
\textbf{Boundary invariant: }Boundary or surface bands may emerge at an interface between two half-spaces made of two bulk systems with a mutual band gap, that is an interval of eigenvalues within which no bulk solution can be found in either domain. The parallel wave vector $\vec{k}_\parallel$ is an element of the surface BZ that depends on lattice inclination and topologically constitutes a $d{-}1$-torus. The projected bulk bandstructure $\Lambda(\vec{k}_\parallel) \eq \cup_i \Lambda_i(\vec{k}_\parallel)\eq \cup_i \left\{ \lambda_i(\vec{k}_\parallel + k_\perp \hat{\vec{n}}):\,k_\perp\in\mathbb{R} \right\}$ ($\hat{\vec{n}}$ is the unit vector normal to the domain wall, $i$ iterates over the sorted eigenvalues) defines the intervals where bulk solutions exist. Conversely, the region $\mathbb{R}\setminus\Lambda(\vec{k}_\parallel)$ describes the union of all surface band gaps. A gap between $\Lambda_i$ and $\Lambda_{i+1}$ is called partial if there is a $\vec{k}_\parallel$ so that $\Lambda_i(\vec{k}_\parallel)\cap\Lambda_{i+1}(\vec{k}_\parallel)\ne\emptyset$, and otherwise total. A topological invariant can be assigned to any closed path $\gamma$ within the surface BZ along which there is a total bandgap: The signed number of crossings of the surface band dispersion with an arbitrary continuous function $f(k_\gamma)$ within the band gap ($f(k_\gamma)\notin \Lambda(k_\gamma)$) is a topological integer $\mathcal{N}$.

The topological Chern bulk-boundary correspondence is the statement that the boundary invariant $\mathcal{N}$ is equal to the difference in gap Chern numbers of the two bulk domains. The gap Chern number is the sum over the Chern numbers of all bands below the bandgap with respect to the unique closed 2D manifold\footnote{Such a manifold is closed due to the fact that $k_\perp\in\mathbb{R}/G_\perp\mathbb{Z}$ with $G_\perp$ the smallest reciprocal lattice vector along $\hat{\vec{n}}$. A conceptual problem arises only if $\hat{\vec{n}}$ is at an irrational angle with respect to the primitive lattice vectors so that $G_\perp$ does not exist. We exclude these lattice inclinations here.} $\mathcal{M}$ that projects onto $\gamma$ via parallel projection onto the surface BZ. This version of the Chern bulk boundary correspondence is a generalization of Hatsugai's original statement \cite{PhysRevLett.71.3697}, and is applied frequently in topological photonics \cite{Lu2014}. Rigorous proofs of this and other bulk-boundary correspondences beyond tight-binding models require non-elementary mathematical concepts (see for example \cite{prodan2016bulk}).

Importantly, within the context of this work, the established Chern bulk boundary correspondence introduced here does not rely on the fact that the parameter space is the reciprocal space of a d-dimensional lattice. It must only contain it such that the projection direction $\hat{\vec{n}}$ is along the reciprocal space direction perpendicular to the surface BZ. Consider thus the extended 3D parameter space $(k_x, k_y, a_g)$ that contains the 2D BZ of the hexagonal lattice. We define $a_g$ as the global geometrical perturbation strength, with $a_{\pm} \eq \pm a_g$ in the two half spaces (labeled $\pm$) right and left of the domain wall. Note that the symmetry in the definition of $a_{\pm}$ is chosen for convenience only. In fact, the following argument is in principle valid for any continuous functional dependence on either side.

\begin{figure}[!t]
    \centering
    \includegraphics[width=\columnwidth]{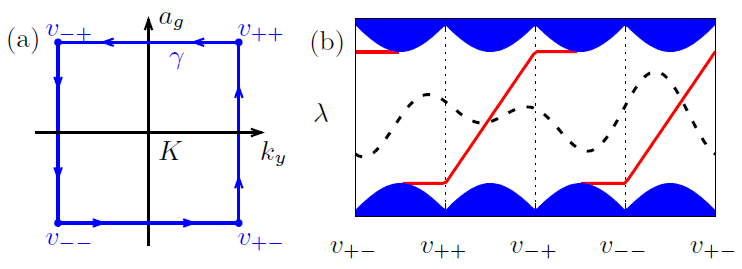}
    \vspace{-1em}
    \caption{Topologically protected surface states in the extended parameter space corresponding to \figref{ky}. (a) The path $\gamma$ encompassing the evaluation point $(K,a_g\eq0)$ in the extended edge BZ parameter space. (b) Along $\gamma$, the signed number of crossings between the topologically protected surface bands (red lines) and an arbitrary continuous curve within the band gap (dashed line) is $\mathcal{N}\eq2$.}
    \label{fig:gamma_BS}
\end{figure}

We have already established in the previous section that the perturbation Hamiltonian is a Weyl Hamiltonian in each spin channel and carries a topological charge of $\pm 1$. Here, the spin channels are separated in $\vec{k}$-space. In other words, any closed surface that contains the Weyl point at say $K$ only will lead to a Chern invariant of $\mathcal{C}\eq\text{sgn}(a)$ for the band immediately below. This non-trivial Chern invariant is equivalent to the gap Chern number as all other bands further below will either be trivial or pair up in form of a similar Weyl point. Let us consider the domain wall along $k_y$ without loss of generality, as illustrated in \figref{ky}. Consider the closed path $\gamma$ in the extended $(k_y,a_g)$ parameter space shown in \figref{gamma_BS} (a). Generally, the corresponding (closed) manifold $\mathcal{M}$ (that projects onto $\gamma$ via parallel projection along $k_x$) sustains a jump in gap Chern number of $\Delta\mathcal{C}_{\text{gap}}\eq2$, and thus predicts that two strongly protected surface bands cross the bandgap with positive slope along $\gamma$. This statement remains correct for any other $\gamma'$ encompassing only $K$, \ie~as long as the inclination is not along the $K$ direction itself and $\delta k_y^0$ is small enough so that $K'$ is outside of $\mathcal{M}$.

\begin{figure}[!t]
    \centering
    \includegraphics[width=\columnwidth]{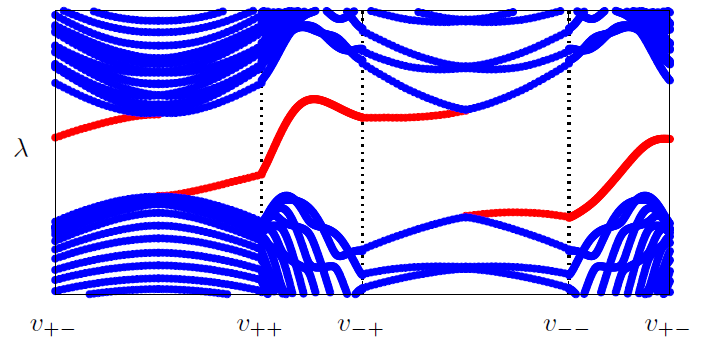}
    \vspace{-2em}
    \caption{Topologically protected surface states along the path $\gamma$ encompassing a finite area in $(k_\parallel,a_g)$ parameter space, \cf~\figref{gamma_BS} (a).}
    \label{fig:gamma_num_BS}
\end{figure} 

In the close vicinity of the evaluation point, we can test the topological prediction analytically, using the same method as in section \ref{sec:QSHE}. Due to the hyperconic dispersion relation, the projected bulk band structure is bounded by the bulk bands for $\delta k_x\eq 0$, \ie~by $\lambda_B\eq\pm\sqrt{a^2+\delta k_y^2}$. Matching the fields on both sides of the domain wall close to $K$ on the other hand yields the surface band dispersion $\lambda_s = -\text{sgn}(a_g) \delta k_y$. This result is shown in \figref{gamma_BS} (b): The predicted boundary invariant $\mathcal{N}\eq 2$ corresponding to $\Delta\mathcal{C}_{\text{gap}}\eq2$ is clearly observed (the red surface dispersion positively crosses the arbitrary dashed gap function twice). For $\gamma$ encompassing a finite area, a simple analytical prediction based on our group theoretical findings is not possible, but the topological protection of the edge states of course still holds, as demonstrated in \figref{gamma_num_BS}. The edge bandstructure is here extracted from a supercell calculation for a 2D photonic crystal based on a perturbed kagome lattice (see \cite{wong2019gapless} for details).

We conclude in noting that no genuine topological protection can be associated with the optical QSHE case, although surface bands that resemble those of proper $\mathbb{Z}_2$ protection can be found for small perturbations in first order perturbation. For the QVHE case, we showed that genuine topological $\mathbb{Z}$ protected edge states exist within the extended parameter space, containing the reciprocal lattice and geometrical perturbations. Within the scope of this manuscript, this protection is only defined within the parameter space of a certain type of a deterministic geometrical perturbation. The above line of thought is, however, applicable to an extended parameter space of any dimension. We note in this context that another pathway to understand topological protection in the QVHE case based on lattice folding at the domain wall has recently been reported \cite{PhysRevB.98.155138}. The requirement imposed in order to be able to map to a topological reference system there is that the two sub-lattices on either side of the domain wall are related by a mirror symmetry. Conceivably, breaking of this mirror symmetry by random perturbations would not allow to define a similar reference system. In contrast, the method presented here is also valid for inclinations where such a mirror symmetry is not possible, and it provides the opportunity to extend the study to a deterministic QVHE system with superimposed arbitrary small random perturbations along other directions in a larger extended parameter space.

\pagebreak
\section*{Acknowledgements}
This work was supported by the EPSRC through program grant EP/L027151/1 and by the European Regional Development Fund through the Welsh Government (80762-CU145 (East)).

\bibliography{./literature}

\appendix

\section{A glossary for representation theory \label{app:glossary}}

We here define and explain the applied concepts of representation theory for finite groups, mainly following \cite{Bradley_GT}. It should be noted that some of the findings from here on are restricted to unitary group elements (expressed by linear operators $\hat{G}$ acting on $\mathcal{H}$), \ie~they satisfy the inner product relation $\langle \hat{G} v|\hat{G} w\rangle = \langle v|w\rangle$ for any any two vectors $|v\rangle,|w\rangle\in\mathcal{H}$. Non-linear symmetries such as hermiticity, reciprocity and time inversion are treated separately. The latter $\hat{\tau}$ is for anti-unitary, with $\langle \hat{\tau} v|\hat{\tau} w\rangle = \cc{\langle v|w\rangle}$ ($\cc{\,\cdot\,}$ denotes complex conjugation), while the other two are fully non-linear and do not have a linear/antilinear operator representation.

To obtain a formally finite group, we replace the translation group $\mathcal{T}$ by the quotient group $\mathcal{T}\mapsto\mathcal{T}/[T]_\approx$, $T\approx T'$ if $\hat{T}'=\hat{T}\prod_i \hat{\vec{a}}_i^{n_i N_i}$, $i=1,2,3$, with $\hat{\vec{a}}_i$ the $i$-th primitive translation, $n_i\in\mathbb{Z}$, and $1\ll N_i\in\mathbb{N}$.  This is a standard simplification technique (\cf~\cite{Roessler}) and equivalent to imposing periodic boundary conditions on a large but finite crystal, or supercell.\\

\noindent\textbf{Representation (rep)}: Consider a homomorphism $\Phi_d: \mathcal{G} \rightarrow \mathcal{D}_{d}$, with $G\mapsto D_d(G)$, that maps any group element $G$ onto a linear $d$-dimensional map $\mathcal{D}_{d}(G):\Omega_d\rightarrow \Omega_d$. We call $\Phi_d$, or equivalently the set of $D_d(G)$, a representation of $\mathcal{G}$, with an associated $d$-dimensional vector space $\Omega_d$.\\

\noindent\textbf{Character}: A representation can be uniquely characterised by the traces of the associated maps $\mathcal{D}(G)$. These traces play a central role in a number of powerful theorems, so that an established synonym for the trace of $\mathcal{D}_d(G)$ is the character of the representation $\mathcal{D}$ with respect to a group element $G$, with symbol $\chi(G)$.\\

\noindent\textbf{Partner}: It is convenient to choose a particular set of basis vectors that span the vector space $\Omega_d$. These basis vectors are also known as partners of the representation. All $D_d(G)$ are then given by unitary square matrices of dimension $d$.\\

\noindent\textbf{Irreducible representation (irrep)}: Consider a vector subspace $V$ of $\Omega_d$. The representation $\mathcal{D}(G)$ is called irreducible, if any such $V$ satisfies the following:
\begin{equation*}
    \forall G\in \mathcal{G}:\;D_d(G) V \subset V \Rightarrow V={0} \lor V=\Omega_d\text{ .}
\end{equation*}
In other words, if we choose a set of partners, there is no similarity transform that brings all matrices $D_d(G)$ into block diagonal form. Irreducible representations can therefore be considered as a generalisation of the concept of simultaneous diagonalization to non-Abelian groups.

The partners of all irreducible representations span the whole associated Hilbert space. This statement is equivalent to the fact that any representation of $\mathcal{G}$ can be written as a direct sum of irreducible representations. Finding the irreps and the partners within the physical Hilbert space is useful in the particular case, because the eigenfunctions of the Hamiltonian $H$ can be identified with a superposition of partners of a single irrep only. The exception to this rule are accidental degeneracies, including exceptional points, which may occur when the eigenvalues of $H$ are \emph{accidentally} (by coincidence for a particular set of parameters of the physical system) glued together. These degeneracies are unpredictable from first principles and rare, we will not consider them in the following. Nevertheless, both cases can be engineered to occur at points or lines in the BZ if the physical properties are desired \cite{Makwana:19,Zhen2015,Huang2011}.

If we thus are able to determine the irreps of a whole space group, we can gain additional insight into the nature of the physical modes, such as prediction of degeneracies and selection rules for field integrals without actual calculation of the eigensystem. Further, we can potentially decrease the dimension of the particular problem for direct computation. For a straight-forward recipe to determine the irreps of $\mathcal{G}$ that are of interest for this work, three additional definitions are essential.\\

\noindent\textbf{Star}: Consider a group $\mathcal{G}$ with an invariant (not necessarily Abelian) subgroup $\mathcal{T}$, \ie~$G\mathcal{T}G^{-1}=\mathcal{T}$ for any $G\in\mathcal{G}$. Denote the irreps of $\mathcal{T}$ by $\Delta_i(T)$. The star of $\mathcal{G}$ with respect to the representation $\Delta_i(T)$ of $\mathcal{T}$ is a maximal set of inequivalent representations $\Delta_i^{(\alpha)}(T):=\Delta_i(G_\alpha^{-1} T G_\alpha)$.\\

\noindent\textbf{Little group}: With the above definitions, the little group of $\mathcal{G}$ with respect to a particular representation $\Delta_i(T)$ is $\mathcal{G}_i :=\left\{ G\in\mathcal{G}: \Delta_i(GTG^{-1})\equiv\Delta_i(T) \right\}$. Note that the little group is evidently a subgroup of $\mathcal{G}$ (albeit not invariant in general).\\

\noindent\textbf{Induced representation}: Consider the expansion of a group $\mathcal{G}$ into left cosets with respect to a subgroup $\mathcal{S}$, that is $\mathcal{G} = \sum_\alpha r_\alpha \mathcal{S}$. Denote the irreps of $\mathcal{S}$ by $\Gamma_j(S)$, where $S\in \mathcal{S}$ and $j$ is an index (not the dimension of the irrep as before). The induced representation of $\Gamma_j$ in $\mathcal{G}$ is written as $\Gamma_j\uparrow\mathcal{G}$. It is most conveniently defined via a set of partners $|n\rangle$, so that the irreps take the matrix form given by $s_a|n\rangle = \Gamma_j^{(nm)}(s_a)|m\rangle$. We introduce a new set of partners $|\mu n\rangle:=r_\mu|n\rangle$, so that the induced representation is given by:
\begin{equation*}
    \Gamma_j\uparrow\mathcal{G}^{(nm)}_{\mu\nu}(G) :=
    \begin{cases}
        \Gamma_j^{(nm)}(G_{\mu\nu}) &\!\text{if } G_{\mu\nu}\!:=\!r_\mu^{-1} G r_\nu \in \mathcal{S} \\ 
        0 & \!\text{else}
    \end{cases}
\end{equation*}
Note that there is a unique pair of $(\mu,n)$ for each space group element $G=r_\mu s_n$, and a (generally different) unique pair $(\nu,m)$, so that the same element is written as $G=s_m r_\nu^{-1}$. Therefore, for a given $\mu$ (or a given $\nu$) there is only one pair $(\mu,\nu)$, for which the induced representation becomes non-zero, and its outer structure is hence that of a permutation matrix. Importantly, we do not mean here that the induced representation can be written as a tensor product $A_{\mu\nu}\otimes B_{nm}$. Given three cosets and an element $G$ that is not in $\mathcal{S}$, it could for example look like:
\begin{equation*}
    \Gamma_j\uparrow\mathcal{G}^{(nm)}_{\mu\nu}(G) =
    \begin{pmatrix}
        0 & \Gamma_j(G_{12}) & 0 \\
        \Gamma_j(G_{21}) & 0 & 0 \\
        0 & 0 & \Gamma_j(G_{33})
    \end{pmatrix}
\end{equation*}
\\
\hspace{9em}
\\
\noindent\textbf{Small representation}: The small representations of a group $\mathcal{G}$ with respect to a subgroup representation $\mathcal{S}$ are those irreducible representations $\Gamma_j^{(i)}$, that contain only a single irrep $\Delta_i$ in $\mathcal{S}$, when restricted to $S\in\mathcal{S}$, \ie~we may write $\Gamma_j(S) = \Delta_i(S) \oplus \Delta_i(S) \oplus \dots$, where $\oplus$ denotes a direct sum.\\

With these definitions, one can show \cite{Bradley_GT} that the irreps of a group $\mathcal{G}$, with an invariant subgroup $\mathcal{T}$ can be obtained by the following steps:
\begin{enumerate}
    \item distribute the reps of $\mathcal{T}$ into stars and select one rep $\Delta_i$ in each star
    \item find the small representations for each associated little group $\Gamma_j^{(i)}$, that contain only $\Delta_i$ in $\mathcal{T}$
    \item the irreps of $\mathcal{G}$ are the induced representations $\Gamma_j^{(i)}\uparrow\mathcal{G}$
\end{enumerate}

\end{document}